\documentclass[11pt,tightenlines,aps,prd,groupedaddress,nofootinbib]{revtex4}

\usepackage{amsmath,amssymb,bm}
\usepackage{graphicx}
\usepackage{epstopdf}
\usepackage{amsfonts}
\usepackage{amssymb}
\usepackage{amsbsy}
\usepackage{amsmath}
\usepackage{latexsym}
\usepackage{bm}
\usepackage{color}
\usepackage{subfigure}
\usepackage{comment}
\usepackage{csquotes}								
\graphicspath{./Images}

\newcommand*{\di}{\partial}

\newcommand{\eq}[1]{(\ref{#1})}

\definecolor{orange}{RGB}{150,0,200}
\newcommand{\ma}[1] {\textcolor{blue}{ #1}}

\def\l{\left}

\def\r{\right}
\def\p {\partial}

\def\be{\begin{equation}}
\def\ee{\end{equation}}
\def\bea {\begin{eqnarray}}
\def\eea {\end{eqnarray}}
\def\nn {\nonumber}

\begin{document}

\title{Mixmaster dynamics in the dust time gauge}

\author{Masooma Ali}
\author{Viqar Husain}

\affiliation{ Department of Mathematics and Statistics, University
of New Brunswick, Fredericton, NB, Canada E3B 5A3} 

\date{\today}

\begin{abstract}

 We study the Hamiltonian dynamics of the dust-Bianchi IX universe in dust time gauge. This model has three physical metric degrees of freedom, with evolution determined by a time-independent physical Hamiltonian.  This approach gives a new physical picture where  dust-Bianchi IX dynamics is described by oscillations between dust-Kasner solutions, rather than between vacuum-Kasner solutions. We derive a generalized transition law between these phases, which has a matter component. Sufficiently close to a singularity, we show that this law reduces to the vacuum Belinski-Khalatnikov-Lifshitz map.  We include an analysis with dust and a scalar field. Lastly, we describe a path integral quantization using the dust-time physical Hamiltonian, and derive an effective action for the dust-Kasner model by  integrating out the anisotropy degrees of freedom.  
 
 \end{abstract}

 \maketitle

\section{Introduction}

In a seminal work Belinski, Khalatnikov and Lifshitz (BKL) studied the approach to spacelike singularities in Einstein gravity by studying the Bianchi IX cosmological model \cite{BKL1,BKL2,BKL3}.   Their analysis led to the so-called BKL conjecture, that the approach to spacelike singularities is universal, with evolution equations dominated only by time derivatives, and characterized by anisotropy oscillations. The conjecture  states that the approach to a spacelike singularity is homogeneous, and therefore it is sufficient to study the most general such models to analyze near singularity dynamics. The literature in this area has since become vast \cite{Berger:1998wr, Berger:1998vxa, Garfinkle:2003bb,Ashtekar:2011ck,Uggla:2013laa}. Recent reviews are \cite{coley-hobill,Garfinkle:2004ww,Berger:2014tev}. These works cover classical aspects of solutions with and without matter, and mini-superspace quantization,which began with Misner's work on Hamiltonian quantum cosmology in \cite{Misner:1969hg,Misner:1969ae}.

The vacuum Bianchi I model is considered the ``free theory"  of anisotropic  cosmology, where dynamics is governed only by the gravitational kinetic term in the Hamiltonian constraint. Its solution is the Kasner metric 
\be
ds^2 = -dt^2 + t^{2p_1} dx^2 +  t^{2p_2} dy^2 + t^{2p_3} dz^2,
\ee
where the the (real) parameters $p_1,p_2$ and $p_3$ are integration constants satisfying the two  sum rules 
\be
p_1+p_2+p_3=1, \ \ \  p_1^2 + p_2^2 +p_3^2 =1. 
\ee
The solution is thus characterized by one free parameter on the so-called Kasner circle at the intersection of this plane and unit sphere.  

More complicated Bianchi models have interactions between the three scale factors arising from the Ricci curvature term in the Hamiltonian constraint.  This is clear in the Hamiltonian formulation of Bianchi IX (or Mixmaster) Universe first studied by Misner  \cite{Misner:1969hg,Misner:1969ae}. The Bianchi IX  potential is an equilateral  triangular box in configuration space with exponentially high sides. The potential vanishes in the region near the origin, so the solution there is the (vacuum) Kasner metric. Bianchi IX dynamics  is thus equivalent to a particle in this box that undergoes collisions at the walls, and after each collision enters a new Kasner phase: 
\be
(p_1,p_2,p_3) \rightarrow (p_1',p_2',p_3').
\ee 

BKL derived a precise transition law for these exponents in the vacuum case, which was subsequently re-derived by Misner in a Hamiltonian formulation cited above.  In the non-vacuum case it is clear that solutions are also labelled by matter integration constants, and so it is natural to expect that these additional constants should also participate in a generalized transition law.  

In this article we derive the following results:
\begin{enumerate}
\item
We formulate Bianchi IX dynamics  in dust time gauge  using  the matter and geometry integration constants for the dust-Bianchi-I solution. We then derive a transition law, akin to the BKL-Misner law, which is valid at all times.
\item
We demonstrate how this general transition law reduces to the BKL-Misner law in the near singularity limit, thus recovering the ``matter does not matter result."
\item
We construct a path integral for a dust-Bianchi I model and integrate out the anisotropies to yield an effective action for the average scale factor. As far as we are aware, this is the first use of matter time gauge for studying the quantum gravity path integral. 

\end{enumerate}

To put our work in perspective, let us highlight the main past works on classical and quantum dynamics of Bianchi IX with matter, which is our main focus. 

There are several analyses of homogenous cosmologies from the covariant and canonical perspectives. The state of work on the canonical side as of the mid-1970s is in the book \cite{Ryan:1975jw}. The book contains an exhaustive analysis of both vacuum and matter coupled models, including canonical quantization. The main developments in this area since then include work by Kuchar and Ryan \cite{Kuchar:1989tj}, who question the validity of minisuperspace quantization by studying one vacuum model embedded in another, 
and Brown and Kuchar  \cite{Brown:1994py}, who analyzed the canonical theory by using a four-component dust field to fix gauge. A simplification of this to a one-component dust is what we use here. 

The recent work  \cite{Ashtekar:2011ck} proposes a Hamiltonian formulation of  the BKL conjecture using new variables;  this uses the vacuum equations in the connection-triad variables.  Also related to our work is an analysis of proper time gauge for the vacuum quantum gravity path integral \cite{Teitelboim:1983fk}; the difference from our approach is that we use matter rather than a geometric variable as a clock. This also leads to the proper-time form of the metric, but with a significantly simpler  physical Hamiltonian. A later work \cite{Berger:1986va} studies the path integral for vacuum Bianchi I cosmology. However, like many other works, the non-vacuum path integral with matter remains  largely unexplored. 

On the covariant side, there are several analyses of the homogeneous Einstein equations  with focus on near-singularity behaviour with dust. Among the pre-BKL papers is \cite{ellis1967dynamics} where pressureless dust cosmologies are studied using a special set of tetrad frames. A generalization of this work to non-zero pressure appears in \cite{ellis:1969}. Post-BKL, a prominent work is  \cite{eardley:1972} which defines and develops the structure of velocity dominated singularities. The lowest  order equations in that work in the decoupling limit, where each space point propagates independently, is equivalent to the dust-Kasner model we study. Its Wheeler-DeWitt quantization was studied in \cite{Liang:1972gz}.  

In addition to these works, more recent application of dynamical systems approach using volume time has produced a large body of results, with a recent compendium appearing in \cite{Wainwright:2005}. Using these methods, the approach to the singularity is analyzed in detail in   \cite{Ringstrom:2000mk}. 

 None of these works, or any others we are aware of, study the generalization of the BKL transition laws using Hamiltonian methods with  matter and using a matter-time gauge. As we will see below, this approach also offers a new physical  picture where the oscillatory dynamics may be viewed as occurring between dust-Kasner phases  in a monotonic time variable. Our analysis demonstrates  that this picture is useful for both the classical analysis of oscillations, and for the quantum theory, where the Wheeler-DeWitt equation does not have to be solved; instead it is replaced by a time-dependent Schrodinger equation (in dust time), albeit with an unusual physical Hamiltonian -- a function of phase space variables which is identical in form to the Hamiltonian constraint, but is not constrained to vanish \cite{Husain:2011tk}.  
 
We present the results stated above from this perspective, beginning in Section II where the dust time gauge in canonical  general relativity is reviewed following \cite{Husain:2011tk}, and slightly generalized, . In Section III we give a new Hamiltonian derivation of the  dust-Bianchi I solution. This is the Heckmann and Sch{\"u}cking \cite{HS} solution\footnote[1]{We became aware of it after this work was completed through Ref. \cite{Khalatnikov:2003ph}}.  In Section IV and V  we give an analysis  of the  approach to the Bianchi IX singularity. We show that there are  anisotropy oscillations between the dust-Bianchi I solutions (which may be viewed as the ``new free theory"), and we derive the corresponding transition rule. The new result here is a general transition law that includes (dust) matter and describes the dynamics of the model at all times. We show that sufficiently close to a singularity, this law reduces to the well-known vacuum BKL map. This recovers the ``matter does not matter" result as a limit of our more general transition rule, which is in accord with analysis using other approaches \cite{Ringstrom:2000mk};  in this sense our results complement these works. In Sec. VI we add an additional   scalar field, and analyze the dust time dynamics using the method of consistent potentials.  In Section VII we develop a path integral quantization in dust-time gauge, and derive an effective action for the average scale factor by integrating out the anisotropy degrees of freedom.  We close in Sec. VIII  with a brief summary of our main results, and an outlook using our dust-time Hamiltonian approach for inhomogeneous models. An appendix contains a detailed outline of related work, which we include for completeness, and for the purpose of comparing our results.

 
\section{The Hamiltonian Theory with dust}
 
We consider GR coupled to dust and any other arbitrary  matter field,
\be
S=\frac{1}{2\pi}\int{d^{4}x\sqrt{-g}R}-\frac{1}{4\pi}\int{d^{4}x\sqrt{-g}\ m(g^{\mu\nu}\di_{\mu}\phi\di_{\nu}\phi+1)}  + \int d^4x\  {\cal L}_M(\chi).
\ee
The  second term is the dust action, and the last is an arbitrary matter Lagrangian. Variation with respect to $m$ gives the condition that the dust field $\phi$ has timelike gradient.

The ADM canonical theory obtained from this action is 
\be
S=\frac{1}{2\pi}\int{dt \ d^{3}x\left(\tilde{\pi}^{ab}\dot{q}_{ab}+p_{\phi}\dot{\phi} + p_\chi \dot{\chi}-N\mathcal{H}-N^{a}\mathcal{C}_{a}\right)},
\label{can-act}
\ee
where the pairs  $(q_{ab},\tilde{\pi}^{ab})$ and $(\phi, p_{\phi})$ are respectively the phase space variables of gravity and dust.  The matter fields are symbolically denoted by $(\chi, p_{\chi})$, although the number of fields and their tensorial structures will depend upon the choice of the matter Lagrangian. The lapse and shift functions,  $N$ and $N^{a}$ are the coefficients of the Hamiltonian and diffeomorphism constraints
\bea
\label{HG}
\mathcal{H} &=&\mathcal{H}^{G}+\mathcal{H}^{D}  + \mathcal{H}^{M},\\
\mathcal{C}_{a}&=&\mathcal{C}^{G}_{a}+\mathcal{C}^{D}_{a} +\mathcal{C}^{M}_{a}\nn\\
 &=&-2D_{b}\tilde{\pi}^{b}_{a}+p_{\phi}\di_{a}\phi + \mathcal{C}^{M}_{a},
\eea
where $\mathcal{H}^G$ is the gravitational part of the Hamiltonian constraint and 
\be
\mathcal{H}^{D}=\frac{1}{2}\left(\frac{p_{\phi}^{2}}{m\sqrt{q}}+m\sqrt{q}(q^{ab}\di_{a}\phi\di_{b}\phi+1)\right).
\ee

The momentum conjugate to the field $m$ is zero since it appears as a Lagrange multiplier in the covariant action. At this point one could enlarge the phase space to treat $m$ and its conjugate momentum as independent degrees of freedom, subsequently eliminating them by gauge fixing. However, it is more straightforward to vary the term $\mathcal{H}^{D}$ in the canonical action with respect to $m$, and use the resulting equation of motion: 
\be
\label{m}
m=\pm\frac{p_{\phi}}{\sqrt{q(q^{ab}\di_{a}\phi\di_{b}\phi+1)}}.
\ee
This can then be substituted back into $\mathcal{H}^{D}$ to give
\be
\mathcal{H}^{D}=  sgn(m)\  p_\phi  \sqrt{q^{ab}\di_{a}\phi\di_{b}\phi+1}, \label{Hd}
\ee
leaving a canonical action for  $(q_{ab},\tilde{\pi}^{ab})$, $(\phi, p_{\phi})$ and the (non-dust) matter phase space variables.
It is readily verified that the constraints remain  first class with this elimination of $m$. We will see in the gauge fixing below how the sign is selected.


\subsection{Dust time gauge}

We now partially reduce the theory  by fixing only a time gauge, and solving the Hamiltonian constraint to obtain a physical Hamiltonian. The spatial coordinates remain unfixed. We use the dust time gauge  \cite{Husain:2011tk,Swiezewski:2013jza} which equates the physical time with the scalar field, i.e., the spatial hypersurfaces are level surfaces of the dust field,
\be
\label{gauge}
\lambda\equiv \phi- \epsilon t \approx 0, \ \ \  \epsilon=\pm 1.
\ee
This is a special case of the Brown-Kuchar matter reference frame that fixes all four coordinate gauges.
The condition (\ref{gauge}) has a nonzero Poisson bracket with the Hamiltonian constraint, so this pair of constraints together is second class. A gauge condition is deemed good if the matrix of Poisson brackets of second class constraints is invertible and demanding that the gauge condition be preserved in time does not lead to new constraints. 

The first of these gives, using (\ref{Hd}), the Dirac matrix of second class constraints 
\be
\label{matrix}
C = \l[ \begin{matrix}  0 & \{\lambda, \mathcal{H}\} \\  \{\mathcal{H}, \lambda \} &0 \end{matrix} \r] =  sgn(m) \l[ \begin{matrix} 0 & 1 \\ -1 & 0 \end{matrix} \r].
\ee
This matrix is invertible everywhere on the manifold. Thus, the dust time gauge does not breakdown at any point and is therefore a robust choice. The second condition, requiring that the gauge condition be preserved in time, gives an equation for the lapse function:
\be
\epsilon = \dot{\phi}= \left. \left\{\phi, \int d^3x \left (N  \mathcal{H}  + N^a \mathcal{C}_{a}\right) \right\} \right|_{\phi=t}  =   sgn(m) N\ . \label{gauge-pres}
\ee

The corresponding physical Hamiltonian density is 
\be
{\cal H}_P = -\epsilon p_\phi =  sgn(m)\ \epsilon \left( \mathcal{H}^{G} + \mathcal{H}^{M} \right) =  N  \left( \mathcal{H}^{G} + \mathcal{H}^{M} \right),
\ee
where the second equality follows from solving the Hamiltonian constraint and the third  using (\ref{gauge-pres}).  We also note that 
the definition of $p_\phi$ from the dust action, in this gauge, gives
\be
p_\phi = \frac{m}{N} \sqrt{q} \dot{\phi} =  sgn(m) \epsilon  \ \frac{  |m|}{N} \sqrt{q} = |m|\sqrt{q} >0.
\ee 
Thus positive physical Hamiltonian for positive dust energy density requires $N=-1$, implying $\epsilon =-1$.   Substituting into    \eq{can-act} gives  the gauge fixed action
\be
S^{GF}=\frac{1}{2\pi}\int{ dt \ d^3x  \left[\tilde{\pi}^{ab}\dot{q}_{ab} +p_\chi\dot{\chi}  +( \mathcal{H}^{G} + \mathcal{H}^{M}) -N^{a}(\mathcal{C}^{G}_{a} +\mathcal{C}^{M}_{a})  \right]},
\label{GF-act}
\ee
up to surface terms, which do not concern us here.  Thus we see that in the dust time gauge the diffeomorphism constraint reduces to that with only the gravity and matter $(\chi,p_\chi)$ contributions, and the physical Hamiltonian is 
\be
H_p = -\frac{1}{2\pi}\int d^3x\, \l(  \mathcal{H}^G + \mathcal{H}^M\r).
\ee
The corresponding spacetime metric is
\be
ds^2 = -dt^2 + (dx^a + N^a dt)(dx^b + N^b dt) q_{ab}. 
\ee 


 \subsection{The spatially homogeneous sector} \label{hom}
 
In the dust time gauge we equate the surfaces of homogeneity with level surfaces of the dust field. The general four dimensional spatially homogeneous metric can then be written as:
 \be
 ds^2 = -dt^2 + q_{ij}(t) \omega^i \omega^j
 \ee
where $\omega^i$ are invariant $1$-forms corresponding to the three dimensional isometry group of the manifold and $N^i = 0$. In the absence of matter fields besides the dust,  the physical Hamiltonian for a spatially homogeneous background is
\be
H_p=-\frac{1}{2\pi}\int d^3x\, \mathcal{H}^G.
\ee
 When $q_{ij}(t)$ is diagonal, a parametrization of the ADM canonical variables is 
 \bea
 q_{ij} &=& \text{diag} [e^{2\alpha_1(t)}, e^{2\alpha_2(t)}, e^{2\alpha_3(t)}],\nn\\
 \pi^{ij} &=& \frac{1}{2}\  \text{diag} [\pi_1(t)e^{-2\alpha_1(t)}, \pi_2(t)e^{-2\alpha_2(t)}, \pi_3(t) e^{-2\alpha_3(t)}], 
 \eea
 so  the canonically conjugate pairs are  $(\alpha_i, \pi_i)$, $i=1,2,3$.  The physical Hamiltonian then takes the form 
 \bea
 \label{hamiltonian}
 H_p &=&  v_0\l[ -\frac{1}{4\sqrt{q}}\l(\frac{1}{2}\sum_i\pi_i^2- \sum_{i<j} \pi_i\pi_j\r) + V({\bf\alpha}) \r] \nn\\
 &\equiv& H_K + V, 
 \eea
 where $V(\alpha)$ is derived from the scalar curvature of the spatial slice, $  \sqrt{q} = \exp\left({\sum_{i=1}^3 \alpha_i}\right)$,  and $v_0$ is a fiducial volume we set to unity. 
 
An alternative set of phase space variables, obtained from the above by canonical transformation, are the Misner variables $\l(\Omega, \beta_+, \beta_-\r)$, and their conjugate momenta (defined in appendix \ref{misner-analysis}). The physical Hamiltonian in these variables is 
\be
\label{hamiltonian-misner}
H_p =  \l[ -\frac{e^{3\Omega}}{24}  \l( p_+^2 + p_-^2 - p_\Omega^2 \r) + V\l(\Omega,\beta_+,\beta_-\r) \r]. 
\ee

We consider here the diagonal Bianchi I and IX spacetimes, for which the potentials  $V\l(\Omega,\beta_+,\beta_-\r)$ are 
\bea
\label{potentials}
V_I &=& 0 \\
V_{IX} &=& -6 e^{-\Omega} \bigg[ \Big[ \frac{2}{3} e^{4 \beta_+} \big(\cosh(4\sqrt{3}\beta_-)-1\big) - \frac{4}{3} e^{-2\beta_+} \cosh(2\sqrt{3}\beta_-) + \frac{1}{3}e^{-8\beta_+} \Big] \bigg], \nn \\
&\equiv & -6e^{-\Omega}\  v(\beta_+, \beta_-).  
\eea

We make use of both sets of variables, the first to give a derivation of the dust-Bianchi I solution, and the second to study Bianchi IX dynamics. In either parametrization, since $H_p$ is a constant of the motion,  the energy density of the dust $m = H_p/\sqrt{q}$ diverges as the metric determinant goes to zero. Thus, $\sqrt{q}\rightarrow 0$ corresponds to a physical singularity.

 \section{Dust-Bianchi I spacetime}
 
 The isometry group of the Bianchi I model is the three parameter group of translations in three dimensional Euclidean space. In the synchronous basis the metric is
  \be
 ds^2= -dt^2 + e^{2\alpha_1(t)} dx^2 + e^{2\alpha_2(t)} dy^2  + e^{2\alpha_3(t)}dz^2. \label{BImetric}
 \ee
The Kasner metric is the vacuum solution of this form. We now derive a metric of the same form with dust, in the dust time gauge. As we will see, this will turn out to be the  Heckmann and Sch{\"u}cking \cite{HS} solution.

The physical Hamiltonian for this model is given by \eq{hamiltonian} with $V(\alpha) = 0$. The Hamilton equations of motion  are  
\begin{eqnarray}
\label{eomdust}
\dot{\alpha}_1 &=& -\frac{1}{\sqrt{q}}\, \l(\pi_1 - \pi_2- \pi_3\r),\quad \{\text{with cyclic perm. on $\pi_i$ for $\dot{\alpha}_2$ and $\dot{\alpha}_3$}\} \nn \\
\dot{\pi}_i &=& H_K .
\end{eqnarray}
The second equation gives
\be
\label{Psol}
\pi_i(t) = H_K t + \lambda_i 
\ee
with integration constants $\lambda_i$.   

\subsection{Kasner solution: $H_K= 0$}
In this case the above evolution equations imply
\be
\left( \sqrt{q}\ \right)\,\dot{} = \frac{1}{4} \l(\pi_1 + \pi_2 + \pi_3\r)
=  \frac{\Lambda}{4}. \label{qdot}
\ee
and  
 \be
\dot{\alpha}_i = \frac{\Lambda-2\lambda_i}{\Lambda t + 4 \delta}, \quad \Lambda = \sum_i \lambda_i.
\ee
 This gives the following solution for the  scale factors $a_i = e^{\alpha_i}$:  
\be
a_i = \xi_i \left(t + \frac{4 \delta}{\Lambda}\right)^{1-\frac{2\lambda_i}{\Lambda}},
\ee
where $\xi_i$ are constants of integration. The scale factors are not all independent, since they satisfy 
\be
\sqrt{q}=a_1a_2a_3= \frac{\Lambda t}{4} + \delta, 
\ee
which is derived from (\ref{qdot}), and $\delta=\sqrt{q}(0)$.

Defining the exponents 
\be
p_i \equiv   {1-\frac{2\lambda_i}{\Lambda}}, \label{pdef}
\ee
we see that $ p_1+p_2 +p_3  = 1$, as for the Kasner solution. Furthermore, substituting the solution (\ref{Psol}) into the physical Hamiltonian $H_K$, and setting $H_K=0$, yields 
\bea
&& \lambda_1^2 + \lambda_2^2  + \lambda_3^2  = 2(\lambda_1 \lambda_2 +   \lambda_1 \lambda_3 +  \lambda_2 \lambda_3 ), \nn\\
&& \implies  p_1^2 +p_2^2 +p_3^2 =1,
\eea
using the definition (\ref{pdef}).   Lastly, we can absorb the integration constants $\xi_i$  in the coordinates, and  redefine $t \rightarrow  t + 4\delta/\Lambda$ to recover the Kasner solution.  Therefore, the dust time gauge, with initial data chosen such that $H_K=0$, gives the vacuum Kasner solution -- an unsurprising result since the dust energy density $m$ vanishes for this case.  We now turn to the   $H_K=$ constant $\ne0$ cases. 

\subsection{Dust-Kasner solution: $H_K >0$} \label{DK}

For $H_K\neq0$ we can invert the expression for the Hamiltonian to obtain an expression for $\sqrt{q}$,
\be
\label{metricdet}
\sqrt{q} = \frac{1}{8}\l(3H_K t^2 + 2\Lambda t + 8 \delta\r).
\ee
This gives 
\be
\label{LSF}
\dot{\alpha}_i =\frac{6H_K\ (H_Kt + \Lambda - 2\lambda_i)}{ \left(3H_K\  t + \Lambda \right)^2+ 24H_K\delta - \Lambda^2} 
\ee
 
$H_K > 0$ requires
\be
\frac{\Lambda^2}{2}> \lambda_1^2 + \lambda_2^2 + \lambda_3^2,
\ee
while the term $24 H_K \delta - \Lambda^2 $ is proportional to 
\be
 \frac{\Lambda^2}{3} - (\lambda_1^2 + \lambda_2^2 + \lambda_3^2).
 \ee
Therefore, for $H_K > 0$, there are two classes of solutions, with initial data satisfying either
\be
\frac{\Lambda^2}{2}> \lambda_1^2 + \lambda_2^2 + \lambda_3^2 > \frac{\Lambda^2}{3},\label{data1}
\ee
or
\be
\frac{\Lambda^2}{3}> \lambda_1^2 + \lambda_2^2 + \lambda_3 ^2. \label{data2}
\ee
At the end of this section we'll show that the second class of solutions is not physically viable.
\\
If \eq{data1} is satisfied, then \eq{LSF} can be integrated  to give
\be
\label{BIsoln}
a_i =\xi_i \l( y- \Gamma\r)^{\frac{1}{3} + \beta_i } \l(y+\Gamma\r)^{\frac{1}{3}-\beta_i},
\ee
where $y = 3H_Kt + \Lambda$, 
\be 
\Gamma^2 = -24 H_K \delta + \Lambda^2, \quad \beta_i = \frac{2}{3\Gamma}\l(\Lambda - 3\lambda_i\r)
\ee
 and $\xi_i$ are integration constants satisfying $\xi_1\xi_2\xi_3 =-1/(24H_K)$.
This  is the Heckmann-Sch{\"u}cking solution 
\be
a_i = \xi_i\, \tau^{p_i} \l(\tau + 2\Gamma\r)^{\frac{2}{3}-p_i},
\ee
as can be seen by defining 
\be
p_i = \frac{1}{3} + \beta_i, \quad  \tau = 3H_K t + \Lambda-\Gamma,
\ee

Interestingly, even for $H_K\ne 0$, the exponents $p_i$ again satisfy
\be
p_1 +p_2 +p_3 = 1, \quad p_1^2 + p_2^2 + p_3^2 =1.
\ee

Addressing now the second set of data  (\ref{data2}),  it is convenient to define $\Gamma^2 \equiv 24H_K\delta - \Lambda^2$. Then the solution of \eq{LSF} 
\be
a_i = \xi_i \l( y^2 + \Gamma^2 \r)^{\frac{2}{3}} \exp \big[ 4B_i \arctan \l( \frac{y}{\Gamma} \r) \big],
\ee
where $B_i = (\Lambda/3 - \lambda_i)/\Gamma$.
Now $\sum_i B_i^2 < 0$, implying that at least one of the $B_i$ is imaginary and the solutions are not physical.

\subsection{Dust-Kasner solution for $H_K< 0$}

For completeness, we also present solutions with $H_K < 0$. These solutions are not physically relevant since they correspond to a negative energy density for the dust.  When $H_K<0$ we have
\be
\frac{\Lambda^2}{3}<\frac{\Lambda^2}{2} < \lambda_1^2 + \lambda^2 + \lambda_3^2.
\ee
This implies that $24H_K\delta - \Lambda^2<0$.  Uo to this change, the solution solution has the same form  
\be
a_i = \xi_i\, \tau^{p_i} \l(\tau + 2\Gamma\r)^{\frac{2}{3}-p_i}
\ee
where $\tau$ and the exponents $p_i$ are defined as before.

\subsection{Dust-Bianchi I spacetime with scalar field}
In the presence of a free scalar field $\chi$ the physical Hamiltonian in the dust time gauge is 
\be
H_p = H_K -    \frac{p_\chi^2}{2\sqrt{q}}.
\ee
The equations of motion for the scale factor remain the same as in \eq{eomdust}. However  equations for the momenta are now
\be
\pi_i = H_p,
\ee
and those for the scalar field are 
 \be
\dot{\chi} = -e^{-\sum_i \alpha_i} p_\chi, \quad \dot{p}_\chi = 0
\ee
Consequently
\be
\pi_{i} = H_pt + \lambda_i. 
\ee
The solution for $\pi_{i}$ substituted into the expression for the physical Hamiltonian $H_p$    gives
 \be
\sqrt{q} = \frac{1}{8}\left( 3H_pt^2  +2\Lambda t +8\delta \right),
\ee
where now 
\be
\delta \equiv \sqrt{q(0)} = -\frac{1}{4H_p} \left(\frac{1}{2} \sum_i \lambda_i^2  -\sum_{i<j} \lambda_i\lambda_j + 2p_\chi^2    \right).
\ee
The solution for the scale factors is 
\be
a_i = \eta_i \left( \tau \right)^{\frac{1}{3} + \beta_i  } \left( \tau + 2\Gamma  \right)^{ \frac{1}{3}-  \beta_i   } 
\ee
where 
\be 
\tau \equiv 3H_p t + \lambda - \Gamma, \ \ \ \Gamma^2 = -24 H_p \delta  + \Lambda^2, \ \ \  
  \beta_i = \frac{2}{3\Gamma} \left(  \Lambda  - 3\lambda_i    \right).  
\ee
(The definition of $\Gamma$ now uses $H_p$ rather than $H_K$ for the pure dust case).  Defining the  exponents
\be
p_i \equiv \frac{1}{3} + \beta_i,
\ee
now gives 
\be
\sum_{i=1}^3 p_i = 1, \quad \sum_{i=1}^3 p_i^2 = 1- \frac{8p_\chi^2}{\Gamma^2}.
\ee 
The second sum rule depends on the value of the conserved scalar field momentum and the integrations constants $\lambda_i$.  This has the correct limits for $p_\chi=0$ (dust only), and for $H_p=0$ (vacuum).

Substituting for the scale factors in the equation of motion for $\chi$ we have 
\be
\dot{\chi} = \frac{8\,p_\chi}{\tau( \tau + 2\Gamma)}, 
\ee
 Since $p_\chi = \text{const}$, this can be integrated to give
\be
\chi = \frac{8\, p_\chi}{2\,\Gamma} \ln \l[\frac{\tau}{\tau+2\Gamma}\r].
\ee

 
 \section{Dust-Bianchi IX spacetime}
 
The  Bianchi IX dynamics is most easily studied from the Hamiltonian perspective using Misner variables. The physical Hamiltonian is 
(\ref{hamiltonian-misner}) with the non-zero potential in (\ref{potentials}).  The metric is 
\be
ds^2 = -dt^2 + e^{-2\Omega} \left( e^{2\beta}\right)_{ij} \omega^i\omega^j  \label{met2}
\ee
where $\omega_i$ are SO(3) covariant 1-forms  and 
\be
\beta_{ij} = {\text {diag}} ( \beta_+ +\sqrt{3} \beta_-,  \beta_+ -\sqrt{3} \beta_-  ,  -2\beta_+).
\ee
The canonical equations of motion  are
\bea
\dot{\Omega} &=& \frac{\exp(3\Omega)}{12}\ p_\Omega, \quad \dot{p}_{\Omega} = - 3 H_p  - 24e^{-\Omega} v(\beta_+,\beta_-)   \\    \label{omegadot}
\dot{\beta}_\pm &=&-\frac{\exp(3\Omega)}{12} \ p_\pm, \quad \dot{p}_\pm = 6 e^{-\Omega}\  \frac{\p v}{\p\beta_\pm}.
\label{betadot}
\eea

\begin{figure}[t]
\begin{minipage}{0.32\textwidth}
  \includegraphics[height = 5cm, width=\linewidth]{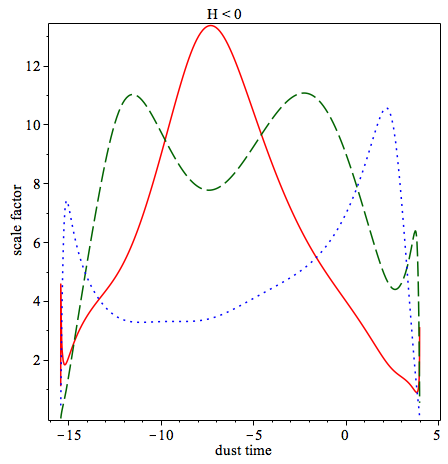}
\end{minipage}\hfill
\begin{minipage}{0.32\textwidth}
  \includegraphics[height = 5cm, width=\linewidth]{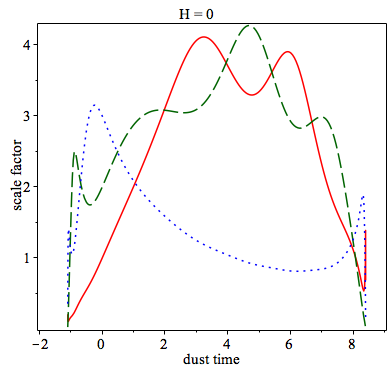}
\end{minipage}\hfill
\begin{minipage}{0.32\textwidth}%
  \includegraphics[height = 5cm, width=\linewidth]{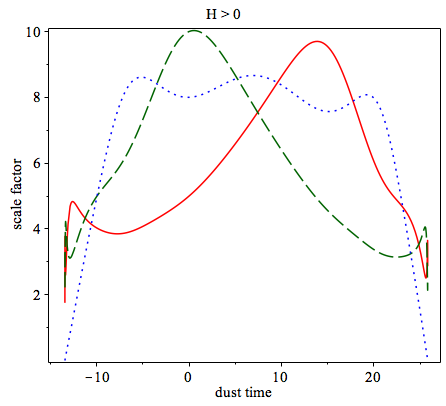}
\end{minipage}
\caption{Numerical integration of the equations of motion for different values of $H_p$ show that oscillatory behaviour of the scale factors $a(t)$ (solid red line), $b(t)$ (dashed green line) and $c(t)$ (dotted blue line).} \label{dynamics}
\end{figure}

The difference between these equations and the vacuum case studied by Misner is that with the dust there are three physical configuration degrees of freedom. Since the dust is used to fix the time gauge, all three degrees of freedom are manifested in the spatial metric, and the potential is a function of all three. Moreover, even though $\Omega$ appears  only in the overall factor, it still has non-trivial dynamics in dust time.

 \begin{figure}[t]
 \centering
\includegraphics[width = 0.5 \textwidth]{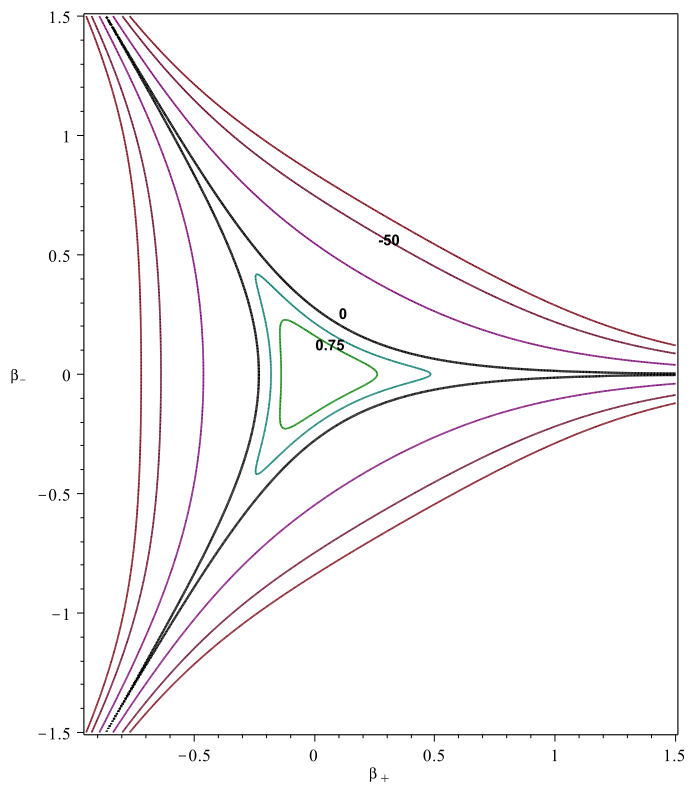}
\vspace{-0.25cm}
\caption{Contours of $v(\beta_+, \beta_-)$ for the potential $V_{IX}= e^{-\Omega}v(\beta_+,\beta_-)$.  $v$ is bounded above by $+1$, the bold line is the $v=0$ contour, the innermost contour is $v=0.75$ and the outermost contour is $v=-100$. } \label{contour-plot}
\end{figure}

The volume of the universe scales as $e^{-3\Omega}$, therefore the singularity is approached as  $\Omega$ tends to infinity.   Thus near the singularity the potential $V_{IX}$  (\ref{potentials}) only plays a role when $|v(\beta_+, \beta_-)|$ is sufficiently large. When the potential term is not dominant, the universe behaves like the dust-Bianchi I system studied in the last section. Therefore, Misner's picture of a particle in a time dependent triangular box can be interchanged with the particle inside a pyramidal well in configuration space depicted in Fig. \ref{contour-plot}. 

Projected on the ($\beta_+,\beta_-$) plane, the contours of $v(\beta_+, \beta_-)$ scale linearly with $\Omega$. As $\Omega$ increases, the contours move outwards. This can be seen by considering one section of the potential, say $V = -2e^{-\Omega - 8\beta_+}$ for 
$\beta_+<0$: setting  $-2e^{-\Omega - 8\beta_+}=-C$  corresponds to contour section given by $ -8\beta_+ = \ln \l(\frac{C}{2}\r) + \Omega$.  The particle velocity in the $(\beta_+,\beta_-)$ plane,  $ \vec{v}=(\dot{\beta}_+,\dot{\beta}_-)$,  scales as $e^{3 \Omega}$ (from the above equations), while the contours have a linear dependence on $\Omega$. It is therefore reasonable to assume that the particle bounces off the exponential walls of the pyramidal potential, and that these bounces are interspersed by durations in which the dynamics is kinetic term dominated and described  by the dust-Bianchi I solution.  

A key difference between the dust time dynamics in the present case, and $\Omega$ time in the standard (no dust) case, is that the singularity is reached in finite dust time. Indeed for $H_p >0$ i.e, the energy density of the dust is positive, the dust filled Bianchi IX universe has two physical singularities as shown in Fig. \ref{dynamics}; this is an example of a more general result \cite{Lin:1990}. 
Fig. \ref{dynamics} also demonstrates that Bianchi IX dynamics in dust time gives oscillatory dynamics, just as for volume time in the vacuum case.

 
  \subsection{Method of Consistent Potentials} \label{mcp}
  
 One way to establish that the Universe particle undergoes bounces at the moving walls of the potential (as the singularity is approached) is a self-consistent analysis called  the Method of Consistent Potentials (MCP) \cite{Grubisic}. The basic idea is to obtain a solution by neglecting the potential terms in the Hamiltonian, ie. a Bianchi I solution, and substitute this solution into the full Hamiltonian, ie.  with the potential terms included. If the dynamics is asymptotically velocity dominated, the neglected potential terms remain exponentially suppressed, i.e. the Bianchi I phase dominates. On the other hand, if one or more of the potential terms grow as the singularity is approached, the Universe may undergo a bounce to a new Bianchi I phase.    
 
 To apply MCP in our case we observe that the physical Hamiltonian for a dust-Bianchi IX spacetime is the sum of two terms $H_K$ and $H_V$ where 
 \bea
 \label{HK}
 H_K &=& - \frac{e^{3\Omega}}{24} \l( p_+^2 + p_-^2 - p_\Omega ^2 \r) \\
 \label{HV}
 H_V &=&  -6 e^{-\Omega} \Big( \frac{2}{3} e^{4 \beta_+} \big(\cosh(4\sqrt{3}\beta_-)-1\big) - \frac{4}{3} e^{-2\beta_+} \cosh(2\sqrt{3}\beta_-) + \frac{1}{3}e^{-8\beta_+} \Big).
 \eea
Near the  singularity, $\Omega \rightarrow \infty$,  so we use the dust-Kasner equations to find $\beta_\pm(\Omega)$ and substitute these into the potential. For large $\Omega$  the Hamilton equations give
\be
\frac{\partial\beta_\pm}{\partial \Omega} = \mp \frac{p_\pm^0}{|L|}  \left(  1- \frac{24He^{-3\Omega}}{L^2}    \right)^{-1/2} 
\approx \mp \frac{p_\pm^0}{|L|} \left( 1+ \frac{12He^{-3\Omega}}{L^2}  \right), 
\ee
where $H$ is the value of the dust-Kasner hamiltonian and $L,p_\pm^0$  are integration constants related by $p_+^0=L \cos \theta$, $p_-^0=L \sin\theta$, a result which follows from the $p_\Omega$ equation. Thus to linear order   we have
\be
\beta_\pm = -\mp \frac{p_\pm^0}{L} \Omega  + \beta_\pm^0.
\ee

Near the singularity the dominant terms in $H_V$ are 
\be 
H_V \approx 2e^{-\Omega} \l(e^{-8\beta_+} + e^{4\beta_+ + 4\sqrt{3}\beta_- } + e^{4\beta_+ - 4\sqrt{3}\beta_ -}  \r),
\ee
which for later convenience we label $V_1$, $V_2$ and $V_3$ respectively. Substituting the asymptotic form of $\beta_\pm$  gives
\be
\label{HVapprox}
H_V \approx 2 \l( e^{\Omega (\pm 4 \sin \theta \pm 4\sqrt{3} \cos\theta  - 1 )} + e^{\Omega(\pm 4 \sin \theta \mp 4\sqrt{3} \cos \theta - 1)} + e^{\Omega(\mp 8 \sin \theta - 1)} \r). 
 \ee
 If all the terms above are to be negligible, we require the following equations to be satisfied simultaneously
 \bea
 \label{conditions}
4 \sin \theta + 4\sqrt{3} \cos \theta -1&<& 0 \nn \\
 4 \sin \theta -4\sqrt{3} \cos \theta -1 &<& 0 \nn \\
 -8 \sin \theta - 1 &<& 0.
 \eea
 
 \begin{figure} [t]
 \centering
\includegraphics[width = 0.5 \textwidth]{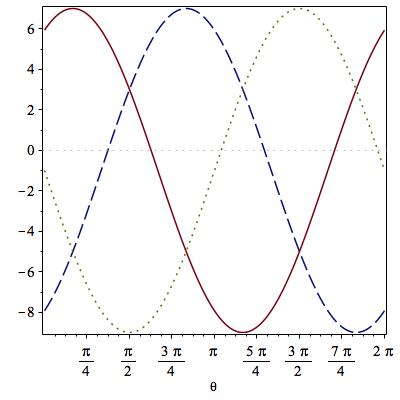} 
\vspace{-0.25cm}
\caption{This is a plot of the expressions on the left hand side of \eq{conditions} with respect to $\theta$. The solid line indicates the first condition, the dashed line indicates the second condition and the dotted line indicates the last condition. It is clear from the plot above that the  inequalities in \eq{conditions} cannot be simultaneously satisfied. Moreover, for any value of $\theta$ only one of the terms in the potential is dominant. }
\label{mcpplot}
\end{figure}

 It is clear from Fig.\ref{mcpplot} that these three conditions cannot be satisfied simultaneously and at least one of the terms is growing at any given time. Thus the particle  is approaching one section of the walls of the pyramidal box at any given time. Therefore the dynamics of the dust-Bianchi IX  near the singularity is characterized by periods in which $H_V$ is negligible compared to $H_K$, and the dynamics resembles that of the dust-Bianchi I model (dust-Kasner phase). These periods are punctuated by periods in which one of the terms in \eq{HVapprox} is large enough that $H_V$ is cannot be neglected causing a \enquote{bounce} from one dust-Bianchi I solution to another. 
 
 {\it Thus, unlike vacuum Bianchi IX, the dust-Bianchi IX universe bounces between Bianchi I solutions that are not vacuum Kasner.} In BKL's language, the dynamics of dust-Bianchi IX is characterised by oscillations between dust-Bianchi I regimes. This gives a new physical  picture of the   approach to the singularity in dust time gauge.

 \section{Transitions between dust-Bianchi I epochs}
  
We have established  that dust doesn't change the oscillatory nature of the Bianchi IX dynamics near the singularity. The dynamics can still be viewed as that of a particle bouncing in a steep triangular potential well, with dust-Kasner regimes between bounces. We would now like to quantify this oscillatory behaviour.

The cornerstone of BKL's analysis of Bianchi IX dynamics is the transition rule governing transitions between various Kasner epochs. In the same spirit we derive a rule that relates the pre- and post-bounce dust-Bianchi I solutions, when these bounces occur away from the corners of the potential.  The method of consistent potentials shows that the three dominant terms in the potential peak at different times. Let us consider first the potential term 
\be
V_1  = -2 e^{-\Omega -8\beta_+}
\ee
which is (a section of) one of the walls of the triangular potential. The truncated Hamiltonian for this wall is then
\be
\label{trunc_ham}
H_1 \equiv  H_K + V_1.
\ee
It is evident that $p_-$ is conserved since the poisson bracket $\{p_-, H_1 \} =0$.   Therefore its change at this potential wall is zero:
\be
\label{transition-cond1}
\Delta p_ - =0.
\ee
However the momentum $p_+$ (which is conserved for dust-Bianchi I)  undergoes a change upon collision. To find this change  let us consider the equations of motion in dust-time: 
\bea
\label{wall1-evol}
\dot{p}_\Omega &=&  - 3 H_1 + 4\ V_1\l(\Omega, \beta_+, \beta_-\r) \\
\dot{p}_+ &=& 8\  V_1\l(\Omega, \beta_+, \beta_-\r).  \label{p+}
\eea
These imply  
\be
p_\Omega - \frac{1}{2} p_+ = -3 H_1 t  + \alpha,  
\ee
where $\alpha$ is an integration constant for the Bianchi IX universe near the section of the potential characterized by $V_1$. Now recalling that $p_+$ and $p_-$ are (approximate) conserved quantities away from the potential wall (where  $H_K\gg V_1$),  the Universe returns to this region with a different value of $p_+$ after a bounce at the wall.  Therefore the dust-Bianchi I regimes before and after collision at the wall $V_{IX} \approx V_1$ are all characterized by the following condition on the integration constants (ie. the last equation evaluated at $t=0$): 
\be
\label{IDcondition}
p_\Omega^0 - \frac{1}{2}p_+^0 = \alpha.
\ee
Now since $\alpha$ is a dust-Bianchi IX integration constant for this wall, we have the relation
\be
\Delta p_\Omega^0 - \frac{1}{2}\Delta p_+^0 = 0, \label{transition-cond2}
\ee
which gives the initial data change for the dust-Kasner phase after collision with $V_1$. {\it This equation is central to our analysis below.}

\subsection{Transition Law: $H_p=0$}
 This is the vacuum case. The following steps give an elegant derivation of the BKL law, which demonstrates the utility of the dust time gauge.      Away from the potential wall we have $H_p \approx H_K=0$, therefore 
\be
p_+^2 + p_-^2 -p_\Omega^2 =0. \label{Kasnercond}
\ee
This suggests the parameterization $\cos \theta \equiv p_+/ p_\Omega$ and $ \sin \theta\equiv p_-/ p_\Omega$. Since $\theta$ undergoes a change at a wall, let us denote  its  values before and after the bounce respectively as  $(p_+/ p_\Omega)^{(i)}  = \cos \theta_i$, $(p_-/ p_\Omega)^{(i)}  = \sin \theta_i$, and   $(p_+/ p_\Omega)^{(f)}  = -\cos \theta_f$, $(p_-/ p_\Omega)^{(f)}  = \sin \theta_f$. ($\theta$ provides an abstract parametrization and in general we cannot interpret it as the angle of incidence or deflection in the $(\beta_+, \beta_-)$ plane.) 

The conservation of $p_-$ at the wall $V_1$ gives 
 \be
 p_\Omega ^{(i)} \sin \theta_i = p_ \Omega ^{(f)} \sin \theta_f,
 \ee 
and   \eq{transition-cond2} gives 
\be
p_\Omega^{(i)} \l(1- \frac{1}{2} \cos \theta_i  \r) = p_\Omega^{(f)} \l(1+\frac{1}{2} \cos \theta_f \r).
\ee
Combining these equations gives 
 \be
\sin \theta_f - \sin \theta_i = \frac{1}{2}\sin (\theta_i + \theta_f). \label{theta-law}
\ee
This rule can be cast in terms of one parameter $u$. Since $p_\Omega$ is a constant when $H_p \approx H_K = 0$, following Misner \cite{Misner:1969ae} we choose the parametrization 
\be
\frac{p_+}{p_\Omega} = \cos \theta=\frac{u^2 + u -1/2}{u^2 + u + 1}, \quad \frac{p_-}{p_\Omega} = \sin \theta= \frac{\sqrt{3}(u + 1/2)}{u^2 + u + 1}. \label{Misneru}
\ee
Then the transition law (\ref{theta-law}) becomes  $u_f = (u_i -1)/3$.

\subsection{Transition Law: $H_p\ne 0$}
 This is the case that gives one of our  new results. It differs from the previous ($H_p = 0$) case in two respects. First, in contrast to (\ref{Kasnercond}), the dust-Kasner physical Hamiltonian (\ref{HK}) now gives
 \be
 p_+ ^2 + p_-^2 =  (p_\Omega^0)^2 - 24H_p \delta  =  \Gamma^2, \label{cond2}
 \ee
 where $\delta =e^{-3\Omega(0)}$  is the initial volume of the dust-Bianchi I solution and $\Gamma$ is defined in Section \ref{DK}. It is important to remember that though the dust-Kasner solution involves six integration constants, the dust-Kasner phase is completely characterized by three integration constants as three constants can be absorbed in redefinitions of the spatial coordinates.
Thus a collision with Bianchi IX wall $V_1$ induces the shift:
 \be
 (p_\Omega^0, p_+, \delta)  \longrightarrow (p_\Omega^{0\ \prime} , p_+', \delta'). 
 \ee
 Importantly, the shift in $\delta$ is now relevant since $ H_p\ne 0$.  Its role is critical for extracting the matter independence of the near singularity transition law we derive below. 
  
Secondly, in the kinetic term dominated region (i.e. $H_1 \approx H_K$), $p_\Omega$ is not a constant but depends linearly on $t$. 
 Thus $\frac{1}{2}p_+ -p_\Omega \neq \text{const}$. Nevertheless  from \eq{IDcondition} we still have  \eq{transition-cond2} as the relation between the integration constants for dust-Bianchi I before and after the bounce at $V_1$, since this condition was derived from the full Bianchi IX equations at wall $V_1$. 
 
Given \eq{cond2}, we define the modified parametrization, before $(i)$ and after $(f)$ the collision, by 
\bea
\left(\frac{p_+}{\Gamma}\right)^{(i)} &=& \cos \theta_ i,\quad  \left(\frac{p_-}{\Gamma}\right)^{(i)} = \sin \theta_ i,\nn\\
\left(\frac{p_+}{\Gamma}\right)^{(f)} &=& -\cos \theta_ f,\quad  \left(\frac{p_-}{\Gamma}\right)^{(f)} = \sin \theta_ f
\eea
Then the conservation of $p_-$ at the bounce gives, as before,
\be
\Gamma^{(i)} \sin \theta_i = \Gamma^{(f)} \sin \theta_f.
\ee
Note, $\theta$ is a redundant parameter introduced for convenience and the shift in $\theta$ is determined by the shift in $\Gamma$ which in turn is governed by the shift in $p_\Omega^0$ and $\delta$.
Now the condition $\frac{1}{2}\Delta p_+^0 =\Delta p_\Omega^0 $,  combined with the last equation, gives  
\be
\label{newlaw}
\left( \frac{p_\Omega^0}{\Gamma} \right)^{(i)} \sin\theta_f  - \left( \frac{p_\Omega^0}{\Gamma} \right)^{(f)} \sin\theta_i = \frac{1}{2}\sin(\theta_i + \theta_f).
\ee
We note that if $H_p=0$ (ie. no dust) this reduces to the BKL-Misner rule. However, as it stands the transition rule is not complete since we have so far not given a prescription for how $\delta$ changes.
The shift in $\delta$ can be obtained by using the dust-Kasner energy conservation, which for the wall $V_1$ gives
\be
- \Delta(p_+^2)  + \Delta(p_\Omega^2) = 24 H_p \Delta \delta \label{delta1}.
\ee
Equation \eq{newlaw} supplemented by \eq{delta1} is one of our main results. We now show how these equations yield the vacuum BKL-Misner rule.

\subsubsection*{\bf Matter does not matter: $\delta$ transition law} 
As it stands  \eq{newlaw} raises the question of compatibility with results from other approaches which establish that the transition rule is matter independent. We now show that sufficiently close to a singularity, the transition rule is such that $\delta\rightarrow 0$. Therefore $\Gamma \rightarrow p_\Omega$, and our new law reduces to BKL-Misner rule.   We demonstrate this for both the initial and final singularity.  As a  byproduct, we  see that our law is the first generalization to include matter, via a matter time gauge, in the  intermediate region where ``matter begins to matter." 

To establish this let us note the following: the transition law (dust-Kasner$)^{(i)} \rightarrow$ (dust-Kasner$)^{(f)}$ at any wall is governed by the dust-Kasner energy conservation equation   
\be
- \Delta(p_+^2) -\Delta(p_-^2) + \Delta(p_\Omega^2) = 24 H_p \Delta \delta, \label{delta2}
\ee
since the total energy $H_p$ of the Bianchi IX solution does not change.  Now the change in $p_+$ and $p_-$ is bounded since $\dot{p}_\pm$ can be positive or negative at different walls.  Therefore close to a singularity, the sign of $\Delta \delta$ is completely determined by $\Delta(p_\Omega^2)$. 

We now establish that $\Delta \delta>0$ during the expansion phase and $\Delta \delta<0$ during the contraction phase. This is sufficient to show that \eq{newlaw} reduces to the vacuum rule sufficiently close to a singularity. We do this by showing that $\Delta(p_\Omega^2)$ accumulates in one direction.

Let us first note that the dust-Kasner evolution  implies
\be
\label{Bianchi I condition}
\dot{\Omega}^{I} < 0, \quad  \ddot{p}_\Omega^{I} =  0.
\ee
Since the sign of $\dot{\Omega}$ is determined by the sign of $p_\Omega$, we have $p_\Omega < 0$ for dust-Kasner evolution. During the expansion phase $\dot{\Omega}^{I}$ and $\dot{\Omega}^{IX}$ have the same sign ($-$ve), and near the singularity in regions where the potential is significant,   (\ref{omegadot}) gives 
\be
\label{rates}
\dot{p}_\Omega^{IX} \ll \dot{p}_\Omega^{I} < 0.
\ee
Thus, after a bounce from the wall $V_1$, $\dot{p}_\Omega^{IX}$ decreases more than it would due to dust-Kasner evolution alone. This extra decrease implies that the shift in the dust-Kasner parameter $p_\Omega^0$ is negative, i.e. $\Delta p_\Omega^0 <0$. Moreover, as the singularity is approached, the inequality in \eq{rates} grows and so does the magnitude of the shift. Since the inequality in \eq{Bianchi I condition} always holds, if the initial conditions for the initial dust-Kasner phase are set such that $\delta^i = 0$, then $p_\Omega^{0\ i} <0$. (This choice of initial conditions is always possible by shifting the dust time origin by $t_0 = (\Gamma^i - \Lambda^i)/3 H_p$.) 

Thus,  $\Delta (p_\Omega^0)^2 >0 $ and increases with each successive bounce, while $\Delta p_\pm^2$ remains bounded. Therefore, $\Delta \delta >0$ in the expanding phase. A similar argument leads to the conclusion that $\Delta \delta <0$ in the contracting phase.
Therefore, as either the past or future  singularity is approached in dust time,  we have 
\be
  \Gamma \rightarrow  p_\Omega,
\ee  
and our transition law \eq{newlaw} reduces to the matter-independent BKL-Misner rule. 

We note that the transition rules at the other walls ($V_2$ \& $V_3$) will in general be different from the  law derived above. This is because though the quantity conserved at these walls can be obtained by suitably rotating \eq{transition-cond1}, it is not possible to obtain the analog of \eq{transition-cond2} in a similar fashion. Therefore, the transition law at the other walls cannot be transformed into the law derived above. This is evident when we note that though the potential term in the Hamiltonian is invariant under rotations of $\frac{2\pi}{3}$, the kinetic term $H_K$ does not have the same symmetry.

Lastly, we can recover the dust-Kasner scale factors before and after a bounce by noting that the integration constants appearing in the transition rule are related to dust-Kasner exponents by
\be
p_1 = \frac{1}{3}(1 - \cos \theta - \sqrt{3} \sin \theta), \,\,\, p_2 = \frac{1}{3}(1 - \cos \theta + \sqrt{3} \sin \theta), \,\,\, p_3 = \frac{1}{3} (1 + 2 \cos \theta).
\ee
  
  \section{Bianchi IX dynamics with dust and scalar field }

It is known that a scalar field suppresses the oscillations between Kasner regimes that characterize the mixmaster dynamics of a generic approach to a singularity. Berger  \cite{Berger:1999} first used MCP to clarify the role of the scalar field. In this section we apply this technique to investigate the near singularity dynamics of a Bianchi IX universe filled with dust and a homogeneous scalar field. In the dust time gauge, the physical Hamiltonian for a Bianchi IX spacetime with dust and a homogeneous scalar field $\chi$ is 
\be
H_p = -\frac{e^{3\Omega}}{24}\l(p_+^2 + p_-^2  - p_{\Omega}^2 +12\, p_{\chi}^2\r) + V_{IX}(\Omega, \beta_+, \beta_-) - e^{-3\Omega} V_{\chi}(\chi),
\ee
where $p_\chi$ denotes the momentum conjugate to the scalar field and $V_\chi(\chi)$ is the scalar field potential. As in the last section, we can again view the Hamiltonian as a sum of two terms $H_K$ and $H_V$ where
\bea
H_K &=& -\frac{e^{3\Omega}}{24}\l(p_+^2 + p_-^2  - p_{\Omega}^2 +12\, p_{\chi}^2 \r) \label{hk}\\
H_V &=& V_{IX}(\Omega, \beta_+, \beta_-) - e^{-3\Omega} V_{\chi}(\chi).
\eea
To apply MCP we are interested in the solutions with the free Hamiltonian $H_K$. The equations of motion are the set (\ref{omegadot})
with $v=0$.  We note also that  from \eq{hk} 
\be
p_+^2 + p_-^2 +12 p_{\chi}^2 = \text{constant} \equiv L.
\ee
Therefore, we can parameterize $p_\pm$ as
\be
p_+ = k\sqrt{L} \sin \theta, \quad p_- = k\sqrt{L} \cos \theta,
\ee
with 
\be
k = \sqrt{1 - \frac{12\, p_\chi^2}{c}}.
\ee
Using the asymptotic expansions for $\beta_\pm$ in $H_V$, near the singularity we have
\be
H_V \approx 2 \l( e^{\Omega (\pm 4k \sin \theta \pm 4\sqrt{3} k\cos\theta  - 1 )} + e^{\Omega(\pm 4k \sin \theta \mp 4\sqrt{3} k\cos \theta - 1)} + e^{\Omega(\mp 8 k\sin \theta - 1)} \r). 
\ee
None of the terms in $H_V$ are significant if the following inequalities are satisfied simultaneously
\bea 
\label{scalarconditions}
\pm 4k \sin \theta \pm 4\sqrt{3} k\cos\theta  - 1 &<& 0 \nn \\
\pm 4k \sin \theta \mp 4\sqrt{3} k\cos\theta  - 1 &<& 0 \nn \\
\mp 8 k\sin \theta - 1 &<& 0.
\eea
All three inequalities are satisfied for $k< 1/4$. Thus, in the presence of the scalar field the oscillatory dynamics of the dust filled Bianchi IX model is suppressed when the scalar field momentum satisfies
\be 
\frac{4}{5} p_\chi^2 > p_+^2 + p_-^2.
\ee
\begin{figure} [t]
 \centering
\includegraphics[width = 0.5 \textwidth]{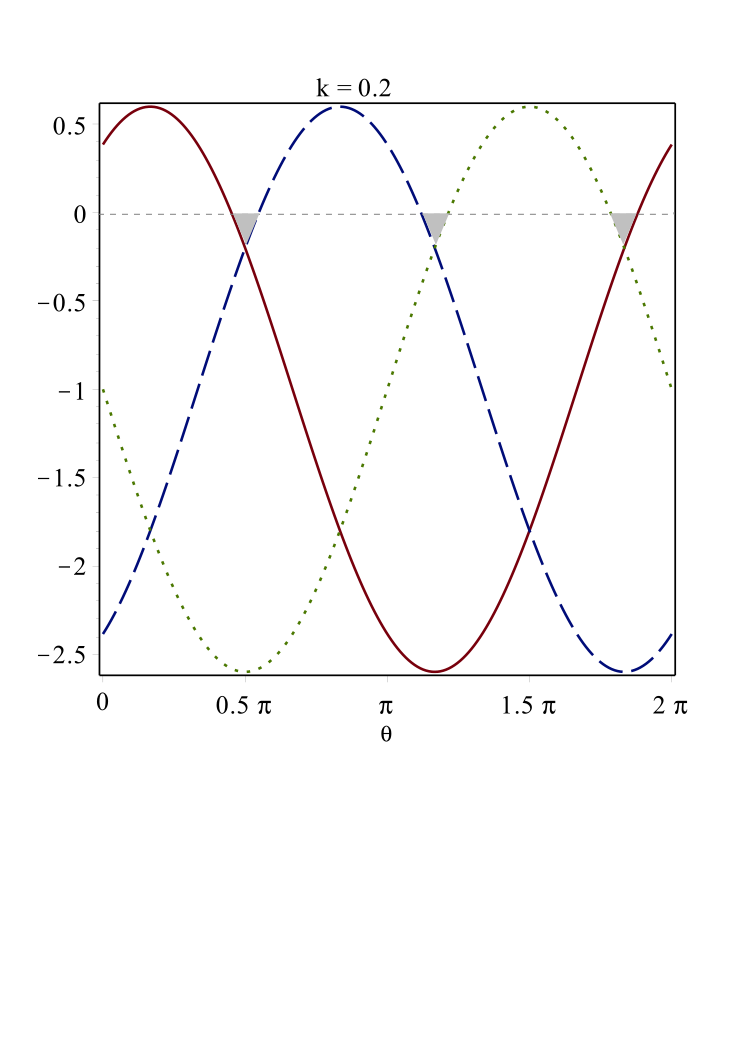}
\vspace{-3.5cm}
\caption{This is a plot of the expressions on the left hand side of \eq{scalarconditions} with respect to $\theta$ for a value of $k<0.25$. The solid line indicates the first condition, the dashed line indicates the second condition and the dotted line indicates the last condition. The gray regions indicate $\theta$ values for which all the terms in the potential are decaying. As the value of $k$ decreases this region grows larger.}
\label{mcpscalar}
\end{figure}

  \section{Dust time path integral}
  
  There is so far no complete quantization of the Bianchi IX system, despite a long history of attempts at solving the Wheeler-DeWitt equation. All attempts involve approximations of one type or another, and are largely qualitative in nature. Here we give a  path integral quantization of the dust-Bianchi I model in the dust time gauge. Specifically we derive an effective action for the scale factor $\Omega$ by integrating  out the anisotropy degrees of freedom $\beta_\pm$  in the  path integral.
     
The canonical action in dust time gauge 
 \be
 S = \int dt \, \left[ \dot{\Omega}\, p_{\Omega} + \dot{\beta}_+ p_+ + \dot{\beta}_- p_- +\frac{e^{3\Omega}}{24} \l(p_+^2 + p_-^2 -p_{\Omega}^2 \r) \right],
 \ee
gives the Lagrangian 
\be
{\cal L} = 6 e^{-3\Omega} \left( \dot{\beta}_-^2  +\dot{\beta}_+^2- \dot{\Omega}^2\right ). \label{lag}
\ee
This resembles a particle version of sigma-model action with non-trivial $\Omega-p_\pm$ interaction. 

An effective action for  $\Omega$ is defined by 
\be
Z[\Omega] \equiv \int \mathcal{D}\beta_- \mathcal{D}\beta_+ \exp \left(\frac{i}{\hbar} \int dt\  6 e^{-3\Omega} \left[ \dot{\Omega}^2 -\dot{\beta}_-^2  -\dot{\beta}_+^2   \right]\right).
\ee
 This is a standard computation. Discretizing time in steps $\epsilon$ and writing $\mathcal{D}\beta = d\beta_1\cdots \beta_{N}$, each  $\mathcal{D}\beta$ integral is 
 \be
 I  = \int d\beta_1\cdots d\beta_{N} \exp\left(\frac{6i}{\hbar \epsilon} \sum_{n=1}^{N} e^{-3\Omega_n}(\beta_n-\beta_{n-1})^2  \right) = \left( \frac{\pi\epsilon \hbar}{6i} \right)^{N/2} \prod_{n=1}^N e^{3\Omega_n/2}.
 \ee
 The full path integral  is then 
 \be
 Z= \int \mathcal{D} \Omega \exp\left( \frac{i}{\hbar} S^{\text{eff}}[\Omega]  \right),
 \ee 
 where the measure is defined to absorb the factors coming from the $\beta_-$ and $\beta_+$ integrations, and 
 \be
 S^{\text{eff}} = \int dt \  \left( 6 e^{-3\Omega}  \dot{\Omega}^2   \right).
 \ee 
The curious feature of this result is that   $S^{\text{eff}}$ is the same as what one would obtain by merely switching  off by hand the anisotropic degrees of freedom. Had we not reabsorbed the $\Omega$ factors coming from the $\beta_\pm$ integrations into the measure, it would not have ben possible to define an effective action in the usual manner. 

The effective equation of motion  is
\be
\ddot{\Omega} - \frac{3}{2} \dot{\Omega}^2 = 0,
\ee
which has the general solution 
 \be
 \Omega = -\frac{2}{3} \ln(c_1 t + c_2). 
 \ee
From \eq{met2} the resulting scalar factor is $a = \exp(-\Omega) = (c_1t+c_2)^{2/3}$, which is the usual matter dominated result. Using this observation, it may be possible to treat the full action for the Bianchi IX case, including the potential, using tunnelling methods between different dust-Kasner vacua.

  \section{Summary and Discussion}
  
 We  studied the Bianchi I and IX cosmologies with dust in the Hamiltonian theory  in the dust time gauge. We first gave a new derivation of the Heckmann-Sch{\"u}cking solution (dust-Kasner) in the dust time gauge, and used this to study the Bianchi IX dynamics.  We showed this approach gives a new physical picture of Bianchi IX evolution,  as a series of dust-Kasner epochs between bounces from the anisotropy potential walls. We then derived the transition law for these dust-Kasner epochs. This law differs  significantly in detail from the vacuum case derived by BKL and Misner, and its form is different at each of the potential walls. 
 
  In the dust time gauge it is not possible to separate the dust degrees of freedom from the gravitational degrees of freedom, since the extra degree of freedom is manifested in the metric, and matter is ``locked in" with time. Therefore it is not surprising that the transition between different dust-Kasner regimes is governed by more than one parameter. 
 
This leads to a puzzle: how does the ``matter does not matter" result arise in a context where evolution is defined with respect to matter time, (dust in our case)? To answer this we showed that the transition rule we derived reduces to the vacuum BKL-Misner law sufficiently close to a singularity.    
  
Lastly we used the dust time classical analysis to develop a path integral quantization with the aim of integrating out the anisotropy degrees of freedom. This is a new idea which bypasses the usual approach to the path integral in quantum gravity, eliminating the need for an integration over the lapse function, and instead using the more tractable physical Hamiltonian in the dust time gauge. For the dust-Kasner case, the effective action for the homogeneous dynamics turned out to be equivalent to  matter-dominated cosmology. The effective action with the anisotropy potential remains open. 
 
The main thrust of the paper is an exploration of Hamiltonian cosmology in the context of the physical time-independent Hamiltonian obtained in the dust time gauge. We consider this to be a potentially useful approach for studying the classical and quantum dynamics of more complex models such as the Gowdy cosmologies. These have so far only been studied to some extent in a volume time gauge, which introduces explicit time dependence in the Hamiltonian and equations of motion. Of particular interest is the approach to the singularity in the quantum theory, which may be simple if the classical equations are indeed vacuum dominated and homogenous.\footnote{D. Garfinkle, personal communication.}  For then the classical singularity to be resolved is  the one provided by a homogeneous cosmology.

\acknowledgments {
This work was supported by the Natural Science and Engineering Research Council of Canada. We thank David Garfinkle for discussions and Edward Wilson-Ewing for helpful comments on the manuscript.}

\appendix
\section{Other approaches}
The study of Bianchi IX dynamics was initiated by Belinskii, Khalatnikov and Lifschitz(BKL) in an effort to characterize the dynamics near a cosmological singularity. The BKL program resulted in the well known BKL conjecture which states that generically the dynamics near a cosmological singularity is vacuum dominated, homogeneous and oscillatory. Bianchi IX spacetimes have since become a subject of interest due to the chaotic behaviour exhibited near the singularity. Various authors have studied different aspects of this chaotic dynamics and the literature on asymptotic dynamics is vast. The techniques used to study this dynamics can be roughly divided into three categories (i) the scattering problem approach, (ii) the particle in a box Hamiltonian approach, and (iii) the dynamical systems method. BKL were the first to treat the dynamics as a scattering problem and this approach is sketched in \ref{BKL-analysis}. Misner introduced the particle in a box approach in \cite{Misner:1969ae}. Misner used ADM variables to formulate vacuum Bianchi IX as a constrained Hamiltonian system and formulated the asymptotic dynamics as the dynamics of a particle in a triangular box. Details of Misner's analysis are given in \ref{misner-analysis}. Several authors have also studied this Hamiltonian system using Ashtekar variables (eg:  \cite{Ashtekar:2011aa},\cite{calzetta1997}).  The third approach to studying Bianchi IX dynamics involves formulating the system in terms of expansion normalized bounded variables to apply dynamical systems techniques. These variables were first introduced in \cite{ellis:1969}. The advantage of the this approach is that it allows one to write the general Einstein equations for homogeneous cosmologies, thus uniting all the Bianchi models within a single framework. Therefore, this approach has been extremely successful in providing rigorous proofs for a variety of hypotheses about the dynamics. However, unlike the other approaches, it does not provide a physical picture for the evolution of a Bianchi IX universe.

\subsection{The BKL Analysis} \label{BKL-analysis}

Belinskii, Khalatnikov and Lifschitz(BKL) were among the first to construct an approximate solution describing the near singularity dynamics of a diagonal vacuum Bianchi IX universe. Here we provide a summary of their analysis.
In a synchronous reference frame the spatial metric for this spacetime is characterised by three scale factors $a(t)$, $b(t)$, $c(t)$  along the spatial directions $l_a$, $m_a$ and $n_a$ ,
\be
\gamma_{ab} = a^2 l_a l_b + b^2 m_a m_b + c^2 n_a n_b.
\ee
The vacuum Einstein's equations for this system are a set of three evolution equations and a constraint for these scale factors (excluding the constraints given by the $R^{0a}=0$ equations):
\bea
\label{EFEs}
\frac{\l(\dot{a}bc\r)\dot{}}{abc} + \frac{1}{2a^2b^2c^2}\l[a^4 - \l(b^2-c^2\r) \r] &=& 0 \nn \\
\frac{\l(a\dot{b}c\r)\dot{}}{abc} + \frac{1}{2a^2b^2c^2}\l[b^4 - \l(c^2-a^2\r) \r] &=& 0 \nn\\
\frac{\l(ab\dot{c}\r)\dot{}}{abc} + \frac{1}{2a^2b^2c^2}\l[c^4 - \l(a^2-b^2\r) \r] &=& 0\nn \\
\frac{\ddot{a}}{a} +\frac{\ddot{b}}{b} + \frac{\ddot{c}}{c} &=& 0,
\eea
where $\dot{f} = \frac{df}{dt}$.
Using the transformations 
\be
a = e^{\alpha}, \quad b = e^\beta, \quad c = e^\gamma, 
\ee
and a new time variable $\tau$
\be
dt = abc\, d\tau,
\ee
\eq{EFEs} take the form
\bea
2\alpha_{\tau\tau} &=& \l(e^{2\beta} - e^{2\gamma} \r)^2 - e^{4\alpha} \nn\\
2\beta_{\tau\tau} &=& \l(e^{2\gamma} - e^{2\alpha} \r)^2 - e^{4\beta} \nn\\
2\gamma_{\tau\tau} &=& \l(e^{2\alpha} - e^{2\beta} \r)^2 - e^{4\gamma}  \label{EFE2}  \\
\frac{1}{2} \l(\alpha + \beta + \gamma \r)_{\tau\tau} &=& \alpha_\tau \beta_\tau + \alpha_\tau \gamma_\tau + \gamma_\tau \beta_\tau,
\eea
with $f_\tau = \frac{df}{d\tau}$.
Assuming there exists a period during the evolution when all terms on the right-hand side of \eq{EFE2} can be neglected, the above equations can be solved exactly to give 
\be
\label{soln1}
a \approx t^{p_1}, \quad b \approx t^{p_2}, \quad c \approx t^{p_3}
\ee
where the exponents satisfy the following conditions
\be
\label{sum_rule}
p_1 + p_2 + p_3 = p_1^2 + p_2^2 + p_3^2 = 1,
\ee
and $t=0$ is a singularity.
BKL called this the Kasner regime. The Kasner regime cannot persist since at least some of the terms on the right-hand side of \eq{EFE2} are growing. Furthermore, in order to satisfy the conditions \eq{sum_rule} the exponents cannot all have the same sign. Without loss of generality we can assume $p_1<0$. Then as the singularity is approached, the Kasner regime is perturbed by the term $e^{4\alpha}$. Neglecting all other terms in \eq{EFE2} we have
\bea
\label{EFEtrunc}
\alpha_{\tau\tau} &=& -\frac{1}{2}e^{4\alpha} \nn\\
\beta_{\tau\tau} &=& \gamma_{\tau\tau}= \frac{1}{2} e^{4\alpha}.
\eea
Thus, according to BKL the Kasner regime defined by \eq{soln1} provides the initial conditions for the evolution characterised by \eq{EFEtrunc}. Essentially BKL treated Bianchi IX dynamics as a scattering problem with the initial and final (asymptotic) states defined by a Kasner regime.
BKL showed that equations (\ref{EFEtrunc}) can be solved exactly to give
\bea
\label{soln2}
a^2 &=& \frac{2|p_1|}{\cosh(2|p_1|\tau)} \nn \\
b^2 &=& b_0^2 \exp\big[2(p_2 - |p_1|)\big] \cosh\l(2|p_1|\tau\r) \nn \\
c^2 &=& c_0^2 \exp\big[2(p_3 - |p_1|)\big] \cosh\l(2|p_1|\tau\r).
\eea
In the limit $t \rightarrow 0$, the asymptotic form of these solutions is identical with \eq{soln1}. In the limit $t\rightarrow \infty$, the asymptotic form of \eq{soln2} is
\be
\label{soln3}
a \approx t^{p'_1}, \quad b \approx t ^{p'_2}, \quad c \approx t^{p'_3}
\ee
with
\bea
t &\approx& \exp\big[ (1+ 2p_1) \tau \big] \nn \\
p'_1 = \frac{|p_1|}{1-2|p_1|}, \quad p'_2 &=& -\frac{2|p_1|-p_2}{1-2|p_1|}, \quad p'_3 = \frac{p_3-2|p_1|}{1-2|p_1|}.
\eea
\eq{soln3} defines a new Kasner regime where $p'_1$ and $p'_3$ are positive and $p'_2$ is the negative exponent. This new Kasner regime is perturbed by the term $e^{4\beta}$ and over time transitions to another Kasner regime. BKL parameterised the Kasner exponents as
\be
p_1(u) = \frac{-u}{1+u+u^2}, \quad p_2(u) = \frac{1+u}{1+u+u^2}, \quad p_3(u) = \frac{u(1+u)}{1+u+u^2}.
\ee
In terms of this parameterisation the transition from one Kasner regime to another can be described as
\bea
\text{if} \quad\quad && p_1(u) < 0 < p_2(u) < p_3(u)\nn \\
\text{then} \quad \quad&& p'_1 =p_2(u-1),  \quad p'_2 = p_1(u-1), \quad p'_3 = p_3(u-1).
\eea
The value of the parameter $u$ decreases by one during each transition. When $u$ becomes less than one, then it transforms as $u \rightarrow 1/u$ since 
\be
p_1\l(\frac{1}{u}\r) = p_1(u), \quad p_2\l(\frac{1}{u}\r) = p_3(u), \quad p_3\l(\frac{1}{u}\r) = p_2(u).
\ee
Further details on these Kasner transitions and the dynamics of vacuum Bianchi IX can be found in the review by BKL \cite{BKL3}.

\subsection{Misner's Analysis of vacuum Bianchi IX dynamics} \label{misner-analysis}

A spatially homogeneous spacetime has (at least) a three dimensional isometry group. These spacetimes can be classified into nine types, as enumerated by Bianchi. For Type A Bianchi models, i.e models with structure constants satisfying $C^i_{ij}=0$, it is possible to arrive at a Hamiltonian formulation by imposing homogeneity at the level of the action. In this paper we consider Bianchi I and IX models, both of which are Type A models. The reduced ADM action after imposing homogeneity is 
\be
S = \frac{1}{2\pi}\int dt\, \omega^1 \wedge \omega^2 \wedge \omega^3   \l(\tilde{\pi}^{ab}\dot{q}_{ab} + p_\chi \dot{\chi} -  N\mathcal{H}\r)
\ee
where the phase space variables only depend on $t$, $\omega^i$ are invariant $1$-forms corresponding to the isometry group of the manifold and the shift function ($N^a$) has been set to zero. The lapse ($N$) function is the coefficient of the Hamiltonian constraint
\be
\mathcal{H} = \mathcal{H}^G + \mathcal{H}^M, \nn
\ee
where
\be
\mathcal{H}^G = -\sqrt{q}R^{(3)} + \frac{1}{\sqrt{q}} \l(\tilde{\pi}^{ab}\tilde{\pi}_{ab} - \frac{1}{2}\tilde{\pi}^2\r).
\ee
The trace of the gravitational momentum is $\tilde{\pi}=q_{ab}\tilde{\pi}^{ab}$, $R^{(3)}$ is the scalar curvature of the spatial hypersurfaces, and $\mathcal{H}^{M}$ is the matter Hamiltonian.   
The corresponding spacetime metric is
\be
ds^2 = -N^2 dt^2 + q_{ij}\omega^i \omega^j.
\ee
 When $q_{ij}(t)$ is diagonal, we can choose 
\bea
 q_{ij}(t) &=& \text{diag} \{a^2(t), b^2(t), c^2(t)\}, \nn \\
 \pi^{ij}(t) &=& \text{diag} \{ \frac{p_a}{2a}, \frac{p_b}{2b}, \frac{p_c}{2c}\}.
 \eea
 The phase space variables are the scale factors and their conjugate momenta $(a(t),p_a(t))$, $(b(t),p_b(t))$ and $(c(t),p_c(t))$. The Hamiltonian constraint takes the form 
 \be
 \label{hamcnstrnt}
 \mathcal{H}^G =  \frac{1}{4abc}[\frac{1}{2}(a^2p_a^2+ b^2p_b^2+c^2p_c^2)-(abp_ap_b + acp_ap_c + bc p_bp_c)] + V(a,b,c) ,
 \ee
 where $V(a,b,c)$ is related to the scalar curvature of the spatial slice. \\
  
A useful set of variables which separate the expansion from the anisotropy were introduced by Misner \cite{Misner:1969hg}. In these variables the spatial metric is parametrized as $q_{ij} = \exp(-2\Omega) (\exp{\beta})_{ij}$, where $\beta_{ij}$ is a symmetric traceless matrix and $e^ {-\Omega}$ is the average scale factor. For the special case when $\beta_{ij}$ is diagonal, we have
\bea
q_{ij} &=& e^{-2\Omega}\text{diag}(2\beta_+ + 2\sqrt{3}\beta_-, 2\beta_+ -2\sqrt{3}\beta_-, -4 \beta_+), \nn \\
\pi^{i}_j &=&\frac{1}{6}\text{diag}\l(p_+ +\sqrt{3}p_-, p_+-\sqrt{3}p_- , -2p_+\r)  - \frac{1}{3} \delta^i_j p_\Omega, 
\eea
where, $\beta_+$ and $\beta_-$ are independent phase space variables with conjugate momenta $p_+$ and $p_-$ respectively. The other phase space variables are $(\Omega, p_\Omega)$. These variables are related to the scale factors and their conjugate momenta by the following canonical transformation
 \bea
 &&\Omega = -\frac{1}{3} \ln\l(abc\r), \quad\,\,\,  p_\Omega = - \l(ap_a + bp_b + cp_c\r)   \label{pO} \nn \\
 &&\beta_+ = \frac{1}{6} \ln\l( \frac{ab}{c^2} \r), \quad\,\,\,  p_+ = ap_a + bp_b - 2cp_c  \label{b+}\nn \\
&&\beta_- = \frac{1}{2\sqrt{3}} \ln \l( \frac{a}{b} \r), \quad  p_- = \sqrt{3} (ap_a -bp_b). \label{b-}
  \eea
The Hamiltonian constraint in these variables is
\be
\label{Misner_H}
\mathcal{H}^G = \frac{e^{3\Omega}}{24}  \l( p_+^2 + p_-^2 - p_\Omega^2 \r) - V\l(\Omega,\beta_+,\beta_-\r).
\ee
\\

Misner carried out a Hamiltonian analysis that paralleled BKL's in \cite{Misner:1969hg}. He proceeded by multiplying the Hamiltonian constraint by  $\sqrt{q}=e^{-3\Omega}$ and reduced the theory by choosing the time gauge $t=\Omega$. The physical Hamiltonian for this choice of time is
\be
H_p = \l[p_+^2 + p_-^2 + e^{-4\Omega}\bigg[ \Big[ \frac{2}{3} e^{4 \beta_+} \big(\cosh(4\sqrt{3}\beta_-)-1\big) - \frac{4}{3} e^{-2\beta_+} \cosh(2\sqrt{3}\beta_-) + \frac{1}{3}e^{-8\beta_+} \Big] \bigg] \r]^{1/2}.
\ee
This Hamiltonian is explicitly time dependent and the system can be viewed as a particle in an expanding triangular box. The singularity is approached as $\Omega \rightarrow \infty$. Therefore, near the singularity the potential term is only dominant when $v(\beta_+, \beta_-) =\Big[ \frac{2}{3} e^{4 \beta_+} \big(\cosh(4\sqrt{3}\beta_-)-1\big) - \frac{4}{3} e^{-2\beta_+} \cosh(2\sqrt{3}\beta_-) + \frac{1}{3}e^{-8\beta_+} \Big]$, is sufficiently large. When the potential term is negligible the Hamiltonian reduces to the Bianchi I (time independent) Hamiltonian 
\be
H_I = \sqrt{p_+^2 + p_-^2},
\ee
and the \enquote{universe particle} behaves as a Bianchi I universe , i.e. it moves along straight lines in the $(\beta_+, \beta_-, \Omega)$ space. Misner parameterized this straight line motion as:
\be
\beta_+ = \Omega \frac{u^2 + u - 1/2}{u^2 + u +1}, \quad \beta_- = \Omega \frac{\sqrt{3}(u + 1/2)}{u^2 + u +1}.
\ee
This parametrization works because $({\beta'}_+)^2 + ({\beta'}_-)^2 = (p_+/H)^2 + (p_-/H)^2 = 1$. Furthermore, during the Bianchi I regime we can also use the alternative parameterization $p_+/H = \cos \theta$ and $p_-/H = \sin \theta$.
\\
The potential walls rise steeply away from $\beta =0$ and the equipotentials form equilateral triangles in the $\beta_+\beta_-$ plane with tunnel like vertices. The central regions of the three equivalent sides are described by the asymptotic forms $V_1 = \frac{1}{3}e^{-8\beta_+}$, $V_2 = \frac{2}{3}e^{4\beta_+ + 4\sqrt{3}\beta_-}$ and $V_3= \frac{2}{3}e^{4\beta_+ - 4\sqrt{3}\beta_-}$. It can be shown that the wall velocity is ${\beta'}_{wall}= 1/2$, thus the universe particle moves twice as fast as the wall in the region $V<1$. Therefore, at finite intervals the trajectory of the particle collides with the potential wall and is deflected from one straight line (Bianchi I) motion to another. This Misner denoted as the bounce.
In order to characterise this bouncing (oscillatory) dynamics we need to relate the parameters of the trajectory after the bounce to those before the bounce. Misner did this by relating the angle $\theta$ after the bounce to that before the bounce. 
Since $\dot{\beta}$ is the velocity of the universe particle, the angle $\theta$ can be interpreted as the angle of incidence of the particle trajectory on to the potential wall. Let the parametrisation of the initial velocities be $(\beta_+')^{(i)} = p_+^{(i)}/H_{(i)} = -\cos \theta_{i}$ and  $(\beta_-')^{(i)} = p_-^{(i)}/H_{(i)} = \sin \theta_{i}$ and the parametrisation after the bounce be  $(\beta_+')^{(f)} = p_+^{(f)}/H_{(f)} = \cos \theta_{f}$ and  $(\beta_-')^{(f)} = p_-^{(f)}/H_{(f)} = \sin \theta_{f}$. The asymptotic form of the Hamiltonian near the wall defined by $V_1$ is
\be
H = \l[ p_+^2 + p_-^2 + \frac{1}{3}e^{-4\Omega -8\beta_+} \r]^{1/2}.
\ee
Then, $p_-$ and the quantity $K = \frac{1}{2}p_+ + H$ are conserved during the bounce. This gives the following two equations
\bea
H_{(i)}\sin \theta_i &=& H_{(f)} \sin \theta_f, \nn \\
H_{(i)}\l(1 + \frac{1}{2}\cos \theta_i \r) &=& H_{(f)} \l(1-\frac{1}{2} \cos \theta_f\r), 
\eea
which lead to the following rule relating $\theta_i$ and $\theta_f$
\be
\sin \theta_f - \sin \theta_i = \frac{1}{2} \sin(\theta_i + \theta_f).
\ee
In terms of the parameter $u$, the above rule is $u_f = u_i -1$.

\subsection{Dynamical Systems Approach} \label{dynamical-sys}
In order to effectively apply dynamical systems techniques to the Bianchi IX system, the Einstein Field Equations must be expressed in terms of bounded variables. Such a set of variables can be obtained in the $1+3$ formalism and are known as expansion normalised variables. In the $1+3$ formalism one begins by choosing an orthonormal frame such that the local line element becomes $ds^2 = \eta_{ab} \sigma^a \sigma^b$ where $\sigma^a$ are dynamics one forms which satisfy the Lie algebra of the symmetry group for some Bianchi type. The time variable ($\tau$) is chosen such that it is constant on the group orbits. The dynamical variables in this approach are the structure constants $\gamma^\alpha_{\mu\nu}$ of the group Lie algebra which are constants on the group orbits but functions of the global time variable. These can be written as
\bea
\gamma^\alpha_{0\beta}  &=& -\sigma^\alpha_\beta - H \delta^\alpha_\beta - \epsilon^\alpha_{\beta\mu}(\omega^\mu + \Omega^\mu) \nn \\
\gamma^0_{0\alpha} &=& \dot{u}_\alpha , \,\, \gamma^0_{\alpha \beta} = -2 \epsilon^\mu_{\alpha\beta} \omega_\mu,
\eea
where $H$ is the expansion scalar, $\sigma_{ab}$ is the shear tensor, $\Omega^a$ is the angular velocity and $\omega^a$ is the vorticity of some fundamental time-like velocity field.
The spatial components can be decomposed as
\be
\gamma^c_{ab} = \epsilon_{abd} n^{cd} + 2a_{[a}\delta^c_{b]}.
\ee 
 The Einstein field equations consist of evolution equations for ($H, \sigma_{ab}, \omega_{a}, n_{ab},a_{a}$) and  four constraint equations. The variables $u_\alpha$ and $\Omega_{a}$ are completely determined by a choice of frame (equivalent to determining the ADM lapse and shift). A key step in the dynamical systems analysis is to normalize all dynamical variables by appropriate powers of the expansion scalar to ensure that the variables remain bounded as the singularity is approached. Using a time coordinate such that the lapse is set to the expansion scalar, $H$ cancels out from the evolution equations for the other varaibles. Thus the state space is reduced to ($ \Sigma_{ab}, \omega_{a}, N_{ab},A_{a}, \Omega$), where each variable is normalized by the expansion scalar and $\Omega$ is the matter density divided by $H^2$. For the class A Bianchi models $A_{a} = 0$, so we neglect this variable from hereon. The evolution can be formulated as
 \be
 \frac{d \mathbb{X}}{d \tau} = f(\mathbb(X)), \,\, \mathbb{X} \in \mathbb{R}^n,
 \ee
 where $\mathbb{X}$ denotes the reduced state space since the evolution of the expansion scalar decouples. For the Bianchi IX models with orthogonal perfect fluid the reduced state space is five dimensional. It is desirable that the state space be compact, that is no variable diverges faster than the expansion scalar as the singularity is approached or goes to zero faster in the late time limit. For all Bianchi models except Bianchi $VII_0$ and $VIII$ the expansion normalized state space is compact. 
 Defining the shear parameter $\Sigma = \frac{\omega^2}{3H^2}$ and the curvature parameter $K = \frac{^{3}R}{3H^2}$ the Friedmanm equation (Hamiltonian constraint) takes the simple form
 \be
 \Sigma^2 + K + \Omega = 1.
 \ee
 The matter degree of freedom satisfies the evolution equation
 \be
 \frac{d\Omega}{d \tau} = [2 q - (3 \gamma -2 )] \Omega,
 \ee
 where $q$ is the deceleration parameter. Thus, $\Omega = 0$ defines an invariant set for this system of equations. This invariant set is also the boundary which defines the evolution of the vacuum Bianchi models. In a seminal paper   \cite{Ringstrom:2000mk}, Ringstrom proved the Bianchi IX attractor theorem that states a generic Bianchi IX orbit has a limit set near past and future infinity which is a subset of the Bianchi IX attractor. The Bianchi IX attractor is the union of the vacuum Bianchi I and Bianchi II invariant sets of the evolution equations for $\mathbb{X}$. The vacuum Bianchi I invariant set is characterized by $N_a = 0$ for $ a = 1..3$ and $\Omega = 0$ and is nothing but the Kasner circle discussed in the metric variable approach. The vacuum Bianchi II subset is characterized by one $N_a > 0$ and the other two equal to zero along with $\Omega = 0$. The Bianchi IX attractor theorem formalizes BKL and Misner's analysis of the near singularity Bianchi IX dynamics in terms of transitions between Kasner solutions via Bianchi II bounces. Ringstrom showed that the $\Omega$ tends to zero in the asymptotic limit and therefore the Bianchi IX attractor is the union of vacuum invariant sets. This is the analogue of $\sqrt{q^0}$ tending to zero as the singularity is approached. Further details about these variables and the use of dynamical systems technique in cosmology refer to \cite{Wainwright:2005}. Details about the Bianchi IX attractor and the proof of theorem can be found in \cite{Ringstrom:2000mk} and \cite{Heinzle:2009eh}. 

\bibliography{dust-bkl}

\end{document}